\documentclass[12pt,preprint]{aastex}
\slugcomment{To appear in the Astrophysical Journal.}
\slugcomment{\bf DRAFT: \today}
\shorttitle{Interstellar N/O}
\shortauthors{Knauth et al.}

\begin{document}

\title{The Interstellar N/O Abundance Ratio: Evidence for Local Infall?\altaffilmark{1}}

\author{David C. Knauth\altaffilmark{2},
David M. Meyer\altaffilmark{2},
and James T. Lauroesch\altaffilmark{2}$^,$\altaffilmark{3}}

\altaffiltext{1}{Based on data obtained by the NASA-CNES-CSA {\it Far Ultraviolet Spectroscopic Explorer (FUSE)} mission operated by the Johns Hopkins University.  Financial support to U.S.  participants has been provided by NASA contract NAS5-32985.}

\altaffiltext{2}{Department of Physics and Astronomy, Northwestern University, Evanston, IL 60208; d-knauth@northwestern.edu; davemeyer@northwestern.edu; jtl@elvis.astro.northwestern.edu.}
\altaffiltext{3}{Current Address: Department of Physics and Astronomy,  University of Louisville, Louisville, KY 40292.}
{}

\vfill
\begin{abstract}
Sensitive measurements of the interstellar gas-phase oxygen abundance have revealed a slight oxygen deficiency ($\sim$ 15\%) toward stars within 500 pc of the Sun as compared to more distant sightlines.  Recent $FUSE$ observations of the interstellar gas-phase nitrogen abundance indicate larger variations, but no trends with distance were reported due to the significant measurement uncertainties for many sightlines.  By considering only the highest quality ($\geq$ 5 $\sigma$) N/O abundance measurements, we find an intriguing trend in the interstellar N/O ratio with distance.   Toward the seven stars within $\sim$ 500 pc of the Sun, the weighted mean N/O ratio is 0.217 $\pm$ 0.011, while for the six stars further away the weighted mean value (N/O = 0.142 $\pm$ 0.008) is curiously consistent with the current Solar value (N/O = 0.138$^{+0.20}_{-0.18}$).  It is difficult to imagine a scenario invoking environmental (e.g., dust depletion, ionization, etc.) variations alone that explains this abundance anomaly.  Is the enhanced nitrogen abundance localized to the Solar neighborhood or evidence of a more widespread phenomenon?  If it is localized, then recent infall of low metallicity gas in the Solar neighborhood may be the best explanation.  Otherwise, the N/O variations may be best explained by large-scale differences in the interstellar mixing processes for AGB stars and Type II supernovae.  
\end{abstract}


\section{Introduction}
The CNO group constitutes the most abundant elements in the Galaxy after hydrogen and helium.  As such, these elements are heavily involved in the life cycles of stars of all masses and compositions (Wheeler, Sneden, \& Truran 1989) and strongly influence the character of the interstellar medium through which these stars recycle material.  Consequently, the current epoch abundances and abundance spreads of the CNO elements are important to establish accurately for studies of Galactic chemical evolution (Timmes, Woosley, \& Weaver 1995) and for comparisons with the abundances of the young galaxies sampled by QSO absorption-line systems (Timmes, Lauroesch, \& Truran 1995).  
Carbon and oxygen are produced during helium shell burning and returned
to the ISM through Type II supernovae, while nitrogen is primarily formed during the CNO
cycle and expelled into the ISM through the massive winds of Asymptotic Giant
Branch (AGB) stars.  The abundance of interstellar carbon and oxygen have been well studied (e.g., Cartledge et al. 2004; Sofia et al. 2004).  However, due to the paucity of accurate interstellar nitrogen abundance measurements, the situation for interstellar nitrogen is not as secure.  

The abundance of interstellar oxygen (Meyer, Jura, \& Cardelli 1998) is relatively constant, O/H$_{\rm{tot}}$ = (3.43 $\pm$ 0.15) $\times$ 10$^{-4}$, where $N$(H$_{\rm{tot}}$)=$N$(\ion{H}{1}) + 2 $N$(H$_2$), with any variability less than the measured 1-$\sigma$ uncertainties for lines of sight reaching $\sim$ 1 Kpc.  Meyer et al. (1998) found no significant variation of O/H with the fractional abundance of H$_2$ [$f$(H$_2$)], a measure of the physical conditions in the gas.  The lack of a change in abundance with change in $f$(H$_2$) supports the notion that depletion onto interstellar grains is not a major sink of \ion{O}{1}.  Their results showed that \ion{O}{1} can be used to trace the \ion{H}{1} column density in diffuse and translucent clouds.   More recent studies of O/H$_{\rm{tot}}$ (e.g., Cartledge et al. 2004) find evidence for slight depletion effects ($\sim$ 0.1 dex) for a few lines of sight and a weak trend of increased O/H$_{\rm{tot}}$ ($\sim$ 0.1 dex) for lines of sight with $d$ $\geq$ 1 Kpc with relatively small scatter about the mean for all measurements.

Using high-quality {\it Hubble Space Telescope (HST)} data on the weak interstellar  \ion{N}{1} $\lambda\lambda$1159.837, 1160.917 doublet toward seven moderately reddened stars ($d$ $\leq$ 500 pc), Meyer, Cardelli, \& Sofia (1997) suggested that the interstellar nitrogen abundance is constant with N/H$_{\rm{tot}}$ = (7.5 $\pm$ 0.4) $\times$ 10$^{-5}$.  In an effort to expand on this work, Knauth et al. (2003) used moderate to high signal-to-noise (S/N) {\it Far Ultraviolet Spectroscopic Explorer (FUSE)} data toward an additional 17 stars ($d$ $\leq$ 2.5 Kpc).  Knauth et al. (2003) revealed a systematic trend of lower N/H$_{\rm{tot}}$ with increasing $N$(H$_{\rm{tot}}$) above 10$^{21}$ cm$^{-2}$, see their Figure 2.  They also compared their  \ion{N}{1} measurements to the \ion{O}{1} $\lambda$1356 abundance toward some of the same lines of sight and found a similar trend of decreasing N/O for increasing $N$(H$_{\rm{tot}}$), but with slightly smaller measurement uncertainties.  Focusing on these weak, optically thin transitions enable more precise relative determinations of $N$(\ion{N}{1}) and $N$(\ion{O}{1}) (Meyer et al. 1997; Knauth et al. 2003; Cartledge et al. 2004) by minimizing saturation effects and eliminating the oscillator strength uncertainties associated with column densities determined from different transitions.  Thus, the large scatter and potential deficiency seen in the \ion{N}{1} abundance (e.g., Knauth et al. 2003) is quite perplexing.  Could the scatter be due to measurement uncertainties, enhanced depletion or ionization of interstellar \ion{N}{1}, or does it represent a real cosmic variance?  Here, we re-examine the highest quality \ion{N}{1} and \ion{O}{1} data in the $FUSE$ and $HST$ data archive and from the literature (Meyer et al. 1997; Knauth et al. 2003).

\section{Observations and Data Reduction}
We present high-quality {\it FUSE} and {\it HST} N/O data toward 13 stars in the Galactic disk.  Eleven of the sightlines studied here were originally published in Meyer et al. (1997) and Knauth et al. (2003).  All \ion{N}{1} data, which have existing $HST$ observations of interstellar \ion{O}{1} $\lambda$1356, are presented in Table~\ref{N_H}. Observations of the weak interstellar \ion{N}{1} $\lambda$1160 doublet toward HD~220057 (observation ID D0640801) were obtained with $FUSE$ on 30 September 2004 and from 9 December 1999 $-$ 14 August 2000 for HD~224151 (observation IDs P1224101-03 and S3040201-04).  The data for HD~220057 were acquired with the star in the medium aperture (4~$\!\!^{\prime\prime}$ x 20~$\!\!^{\prime\prime}$; MDRS) and the large aperture (30~$\!\!^{\prime\prime}$ x 30~$\!\!^{\prime\prime}$; LWRS) was utilized for HD~224151.  The data cover the wavelength range 905 $-$ 1185 \AA\ with a spectral resolution equivalent to $\Delta$$v$ $\sim$ 20 km s$^{-1}$.  The weak interstellar \ion{N}{1} doublet at 1160 \AA\ appears in both the LiF1 and LiF2 channels which provides a 
consistency check to rule out the possibility of detector artifacts.  We refer the reader to Moos et al. 
(2000) and Sahnow et al. (2000) for a more detailed discussion of the {\it FUSE} instrument and its on-orbit performance.

The histogram data were reduced and calibrated with the most recent version of 
CalFUSE\footnote{The CalFUSE Pipeline Reference Guide is available at 
http://fuse.pha.jhu.edu/analysis/pipeline\_reference.html.}.  CalFUSE provides the appropriate Doppler corrections to remove the effects of the spacecraft motion and places the data on the 
heliocentric velocity scale. The wavelength solution provides 
good relative calibration across the LiF channels and CalFUSE v3.1.3 provides the most accurate wavelength calibration of $FUSE$ data to date (e.g., Dixon, Sankrit, \& Otte 2006, in press).  In order to minimize the uncertainties in the relative wavelength calibration between exposures, the 
data were co-added with a cross correlation technique (Friedman et al. 2006).   The 
final summed spectra (near 1160 \AA) have S/N ratios $\sim$ 100 per pixel for both data sets.   

\section{Results and Analysis}
Knauth et al. (2003) revealed an interesting trend of decreasing nitrogen abundance with increasing hydrogen column density, but with no apparent trends with distance due to the large uncertainties in some of the measurements.   However, when one focuses on only the highest quality data ($\geq$ 5-$\sigma$),  as shown in Figure~1, there is a clear break in the data at a distance of 500 pc with stars closer (solid squares) having N/O = 0.217 $\pm$ 0.011, which is over 50\% larger than N/O = 0.142 $\pm$ 0.008  for stars at further distances.   The 500 pc distance is comparable to the distance (500-800 pc) for which the \ion{O}{1} abundance is lower than expected (Andr\'{e} et al. 2003; Cartledge et al. 2004).   Although the $N$(O~{\small I}) abundance is $\sim$ 15\% lower ($\sim$ 0.1 dex) for stars closer than $\sim$ 500~pc, this oxygen trend alone cannot explain the large ($\sim$ 0.3 dex) disparity in the N/O ratio between local and more distant lines of sight.  It is interesting to note that the current Solar value N/O = 0.138$^{+0.20}_{-0.18}$ (thin solid line; Lodders 2003) is in good agreement with the interstellar value for stars with $d$~$\geq$~500 pc.  The decrease in N/O beyond 500 pc strongly suggests enhanced nitrogen production in the Solar neighborhood or large-scale differences in interstellar mixing processes.  A similar bi-modal trend in N/O observed in Galactic and extragalactic H~{\small II} regions is explained by differences in the production of nitrogen at different metallicities (Henry, Edmunds, \& K\"{o}ppen 2000).  Yet, is it possible that the observed decline in the N/O abundance ratio could be the result of saturation, ionization, or depletion effects?

Saturation of absorption lines due to unresolved velocity structure result in lower abundance measurements from low-to-medium resolution data (Savage \& Sembach 1991) and are potentially an issue for {\it FUSE} observations of the weak interstellar \ion{N}{1} doublet.   For the lines of sight presented here, we find that both members of the \ion{N}{1}  $\lambda$$\lambda$1160, 1161 doublet, which have oscillator strengths of $f_{1159}$ = 9.95 $\times$ 10$^{-6}$ and $f_{1160}$ = 2.75 $\times$ 10$^{-6}$ (Morton 2003), yield the same nitrogen column density within the measured 1-$\sigma$ uncertainties.  In order to further address potential saturation effects, we compare column density measurements of \ion{N}{1} to that of \ion{O}{1} $\lambda$1356 ($f_{1356}$ = 1.16 $\times$ 10$^{-6}$; Morton (2003)) because they have similar $f$-values.  Previous work on high-resolution {\it HST} observations ($\Delta$$v$ $\leq$ 3.5 km s$^{-1}$)  of interstellar \ion{O}{1} $\lambda$1356 (Meyer et al. 1998; Andr\'{e} et al. 2003; Cartledge et al. 2004) find this line to be optically thin with a few exceptions and therefore, generally unsaturated.   The ratios of $Nf\lambda$ for these three transitions are $Nf\lambda_{1160}$/$Nf\lambda_{1356}$ = 1.6 and $Nf\lambda_{1160}$/$Nf\lambda_{1356}$ = 0.44, which implies the weakest member of the \ion{N}{1} doublet is optically thin.  Since both \ion{N}{1} lines yield the same column density, saturation effects are smaller than the measurement uncertainties and therefore negligible.  Furthermore, by comparing N to O through the N/O abundance ratio, the impact of any conceivable {\it uncorrected} saturation for these similarly weak \ion{N}{1} and \ion{O}{1} lines should divide out.
 
The enhanced ionization of interstellar \ion{N}{1} (IP=14.5 eV) over that of interstellar \ion{O}{1} (IP=13.6 eV) has been shown to occur in diffuse interstellar clouds with  $N$(H$_{\rm{tot}}$) $\ll$ 10$^{19}$ cm$^{-2}$ (Jenkins et al. 2000) because \ion{O}{1} is more strongly coupled to \ion{H}{1} through charge-exchange reactions (Sofia \& Jenkins 1998) than is \ion{N}{1}.  Therefore, ionization effects should be negligible toward our  relatively dense diffuse clouds ($N$(H$_{\rm{tot}}$) $\geq$ 10$^{20}$ cm$^{-2}$).  However, ionization, if present, would most likely effect the lower column density (i.e., more local) lines of sight than the denser (i.e., more distant) ones.  As can be seen in Figure 1, our results reveal the opposite trend indicating that ionization effects are not important for these sightlines.

For both N and O, the depletion effects, if any, should be insignificant.  The depletion of interstellar \ion{N}{1} from the gas phase into dust grains is unlikely to be a significant effect for diffuse clouds (Sofia, Cardelli, \& Savage 1994; Sofia 2004).  Cartledge et al. (2001) show small depletion effects ($\leq$ 0.1 dex) in interstellar O~{\small I} for log $N$(H$_{\rm{tot}}$) $\geq$ 21.4 that may only affect 4 lines of sight in our present sample, only two of which are at d $\geq$ 500 pc.  However, given the observed trends in interstellar N/O and the earlier results of Knauth et al. (2003) and Cartledge et al. (2004), depletion effects are unlikely to simultaneously explain the N/O abundance variations.  Therefore, our measured abundance variations imply evidence for a real cosmic variance.

\section{Discussion}
One possible mechanism for cosmic variance arises due to differences in 
interstellar mixing processes.  Nitrogen is primarily formed during the CNO
cycle and expelled into the interstellar medium (ISM) through the winds of intermediate mass (4-8 M$_{\odot}$) AGB stars (Henry et al. 2000), while oxygen is produced during helium shell burning and returned to the ISM through Type II  supernovae.  A larger scatter in the N/H or N/O ratio compared to O/H, would suggest that the products of AGB stars are enhanced in the Solar neighborhood and/or are not as well mixed compared to products of supernovae.  It is generally acknowledged that interstellar mixing processes produce a homogeneous ISM on timescales of $\geq$ 10$^8$ years (e.g., Moos et al. 2002; de Avillez 2000), which suggests that the nitrogen enhancement in the Solar neighborhood is relatively recent.   Additional evidence for enhanced s-process elements in the Solar neighborhood comes from the possible enrichment of Sn with respect to the Solar abundance (Sofia, Meyer, \& Cardelli 1999; Lauroesch et al. 2006, submitted).
Additionally, recent work on the $^{85}$Rb/$^{87}$Rb isotope ratio toward $\rho$~Oph A (Federman, Knauth, \& Lambert 2004) suggests a dearth in r-process Rb in the Solar neighborhood but their findings could represent an enhancement in s-process Rb.  These results all point to a nucleosynthetic origin for the larger N/O abundance due to contributions from AGB stars in the Solar neighborhood.  However, Cartledge, Meyer, \& Lauroesch  (2003) find enhanced krypton abundances toward two distant stars HD~116852 \& HD~152590, which they attribute to additional contributions from low and intermediate mass stars (e.g., AGB stars), hinting at possibly a more wide spread phenomenon.  
   
If localized to the Solar neighborhood, another mechanism could be the infall of low-metallicity gas previously invoked to explain the slight \ion{O}{1} deficiency within a 500-800 pc radius of the Sun (Meyer et al. 1994; Cartledge et al. 2004).  Comer\'{o}n \& Torra (1994) investigated the infall of a low metallicity 10$^6$ M$_{\odot}$ cloud, which explain the dynamics of Gould's Belt.  A recent study on the effect of infalling low metallicity gas on a galaxy (K\"{o}ppen \& Hensler 2005) showed that such an event would have a propensity to form intermediate mass stars.  The infalling material would initally decrease the N/O ratio in a localized portion of the galaxy and then after a time delay of $\sim$10$^8$ years, the timescale needed for AGB stars to begin enriching the ISM, the N/O ratio would increase.  Ultimately, the N/O ratio would return to its canonical value for the host galaxy.  In this scenario, K\"{o}ppen \& Hensler (2005) predict that N/O will be high for low O/H$_{\rm{tot}}$ and low for high O/H$_{\rm{tot}}$.  The right panel of Figure~1 shows that for distances closer than 500 pc, the N/O ratio (solid squares) is higher at lower O/H$_{\rm{tot}}$ than the N/O ratio at further distances (solid circles) indicating that local infall of low-metallicity gas may be the likely explanation (K\"{o}ppen \& Hensler 2005).  Both plots in Figure~1 provide the best evidence that the N/O variability is primarily due to differences in the nucleosynthetic origin of nitrogen and oxygen and not to observational uncertainties.  However, it is important to note that the stellar abundance yields (van den Hoek \& Groenewegen 1997) that K\"{o}ppen \& Hensler (2005) utilized in their models of galactic chemical evolution do not include stellar rotation, which may alter their N/O abundance ratio predictions. 

Is the enhanced N/O ratio in the Solar neighborhood evidence for infall of low-metallicity gas or for large-scale differences in interstellar mixing processes between stellar winds of AGB stars or Type II supernovae?  Only through further precise observations of the interstellar N/O ratio toward stars at variety of distances from the Sun will this question be resolved.  If infall is the solution, then all high N/O ratios should be clustered within a $\sim$ 500 pc radius of the Sun, otherwise further observations will show significant scatter in the N/O ratio at all distances.  If verified, the infall of near-primordial gas in the vicinity of the Sun has important implications for a wide variety of astrophysical problems, including the enhanced D/H ratio within the Local Bubble (Hoopes et al. 2003; Moos et al. 2002; Wood et al. 2004).
 
\acknowledgments   This research has made use of the SIMBAD database, operated at CDS in Strasbourg, France.

\newpage
\begin{deluxetable}{ccccccccc}
\tabletypesize{\scriptsize}
\setlength{\tabcolsep}{0.05in} 
\tablecolumns{7}
\tablewidth{0pt}
\tablecaption{Interstellar N/O Ratios$^1$ \label{N_H} } 
\tablehead{ 
\colhead{Star} & \colhead{distance} &  
 \colhead{$N$(H$_{\rm{tot}}$)} & \colhead{$N$(N I)} & 
\colhead{$N$(O I)} & \colhead{N/O} & \colhead{Ref} \\
\colhead{}  & \colhead{[pc]}  &\colhead{[cm$^{-2}$]} &
\colhead{[cm$^{-2}$]} & \colhead{[cm$^{-2}$]} & \colhead{} & \colhead{}  }
\startdata
$\delta$~Sco & 123$^{+15}_{-12}$ & (1.45 $\pm$ 0.47) $\times$ 10$^{21}$ & (1.17 $\pm$ 0.06) $\times$ 10$^{17}$ 
 & (5.3 $\pm$ 0.6) $\times$ 10$^{17}$ 
 & 0.22 $\pm$ 0.03 & 2, 3 \\
$\zeta$~Oph  & 140$^{+16}_{-13}$ & (1.38 $\pm$ 0.38) $\times$ 10$^{21}$ & (1.05 $\pm$ 0.13) $\times$ 10$^{17}$  & (4.3 $\pm$ 0.4) $\times$ 10$^{17}$
 & 0.24 $\pm$ 0.04 & 3, 4, 5, 6 \\
$\gamma$~Cas & 188$^{+22}_{-18}$ & (1.45 $\pm$ 0.27) $\times$ 10$^{20}$ & (1.15 $\pm$ 0.11) $\times$ 10$^{16}$ & (5.8 $\pm$ 0.5) $\times$ 10$^{16}$ 
 & 0.20 $\pm$ 0.03 & 3, 4, 5, 6 \\
$\kappa$~Ori & 221$^{+46}_{-32}$ & (3.39 $\pm$ 0.31) $\times$ 10$^{20}$ & (2.54 $\pm$ 0.14) $\times$ 10$^{16}$ & (1.1 $\pm$ 0.1) $\times$ 10$^{17}$ 
 & 0.23 $\pm$ 0.03 & 3, 4, 5, 6 \\
$\lambda$~Ori  & 324$^{+109}_{-66}$ & (6.46 $\pm$ 1.04) $\times$ 10$^{20}$ & (5.15 $\pm$ 0.30) $\times$ 10$^{16}$ &  (2.1 $\pm$ 0.3) $\times$ 10$^{17}$ 
 & 0.25 $\pm$ 0.04 & 3, 4, 5, 6  \\
 HD~220057 & 440$^{+166}_{-100}$ & (1.86 $\pm$ 0.33) $\times$ 10$^{21}$ & (1.09 $\pm$ 0.08) $\times$ 10$^{17}$ & (6.17 $^{+0.91}_{-0.80}$) $\times$ 10$^{17}$
 & 0.18 $\pm$ 0.03 & 7, 8 \\
 HD~99857  & 483$^{+2017}_{-216}$ &  (2.51 $\pm$ 0.23) $\times$ 10$^{21}$ & (1.78 $\pm$ 0.24) $\times$ 10$^{17}$ &  (7.8 $\pm$ 0.5) $\times$ 10$^{17}$
 & 0.23 $\pm$ 0.03 & 9, 10 \\
HD~210839 & 505$^{+153}_{-95}$ & (2.82 $\pm$ 0.32) $\times$ 10$^{21}$ & (1.84 $\pm$ 0.19) $\times$ 10$^{17}$ & (1.30 $\pm$ 0.06) $\times$ 10$^{18}$ 
 & 0.14 $\pm$ 0.02 & 9, 10  \\
HD~195965 & 524$^{+228}_{-122}$  & (1.09 $\pm$ 0.13) $\times$ 10$^{21}$ & (8.36 $\pm$ 0.84) $\times$ 10$^{16}$ & (5.89~$^{+~0.57}_{-~0.76}$)
 $\times$ 10$^{17}$ & 0.14 $\pm$ 0.02 & 9, 11 \\
 HD~124314 &  708$^{+1513}_{-287}$ & (3.18 $\pm$ 0,40) $\times$ 10$^{21}$ & (2.21 $\pm$ 0.27) $\times$ 10$^{17}$  & (1.53 $\pm$ 0.08) $\times$ 10$^{18}$ 
 & 0.14 $\pm$ 0.02 & 9, 10 \\
HD~24534 &  826$^{+2877}_{-361}$ & (2.19 $\pm$ 0.38) $\times$ 10$^{21}$ & (1.10 $\pm$ 0.20) $\times$ 10$^{17}$ & (7.43 $\pm$ 0.28) $\times$ 
 10$^{17}$ & 0.15 $\pm$ 0.03 & 9 \\
HD~224151  & 1360 & (2.91 $\pm$ 0.38) $\times$ 10$^{21}$ & (1.56 $\pm$ 0.10) $\times$ 10$^{17}$ 
 &   (1.14 $\pm$ 0.08) $\times$ 10$^{18}$ 
 &  0.13 $\pm$ 0.02 & 7, 10  \\
HD~218915 & 2083  & (1.61 $\pm$ 0.24) $\times$ 10$^{21}$ & (1.06 $\pm$ 0.22) $\times$ 10$^{17}$ 
 & (6.6 $\pm$ 0.4) $\times$ 10$^{17}$ 
 & 0.16 $\pm$ 0.03 & 8, 9, 10 \\
\enddata
\begin{center}
\footnotesize{References$-$ (1) In order of increasing distance from the Sun using Hipparcos parallax
(Perryman et al. 1997); (2) Keenan, Hibbert, \& Dufton (1985);
(3) Meyer et al. (1997); (4) Meyer et al. (1998) with updated O I $\lambda$1356
$f$-value (Welty et al. 1999); (5) H$_2$ from Savage et al. (1977) and Rachford et al. (2002);
(6) H I weighted mean of Bohlin, Savage, \& Drake (1978) and Diplas \& Savage
(1994); (7) This Work; (8) Cartledge et al. (2004); (9) Knauth et al. (2003); (10) Andr\'{e} et al. (2003);
(11) Hoopes et al. (2003)}
\end{center} 
\end{deluxetable}

\begin{figure}[ht]
\vspace{-0.4in}
\epsscale{0.85}
\plottwo{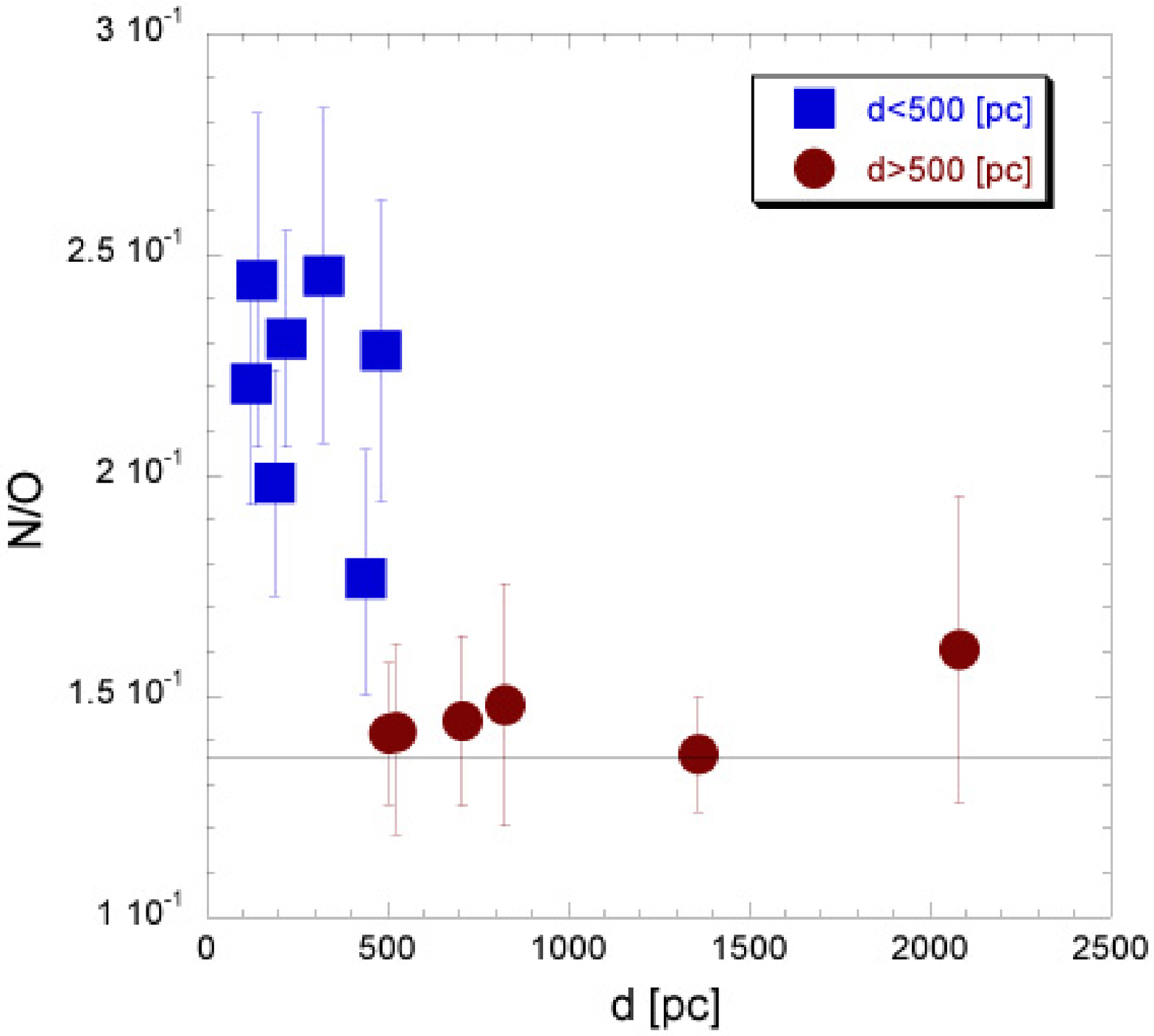}{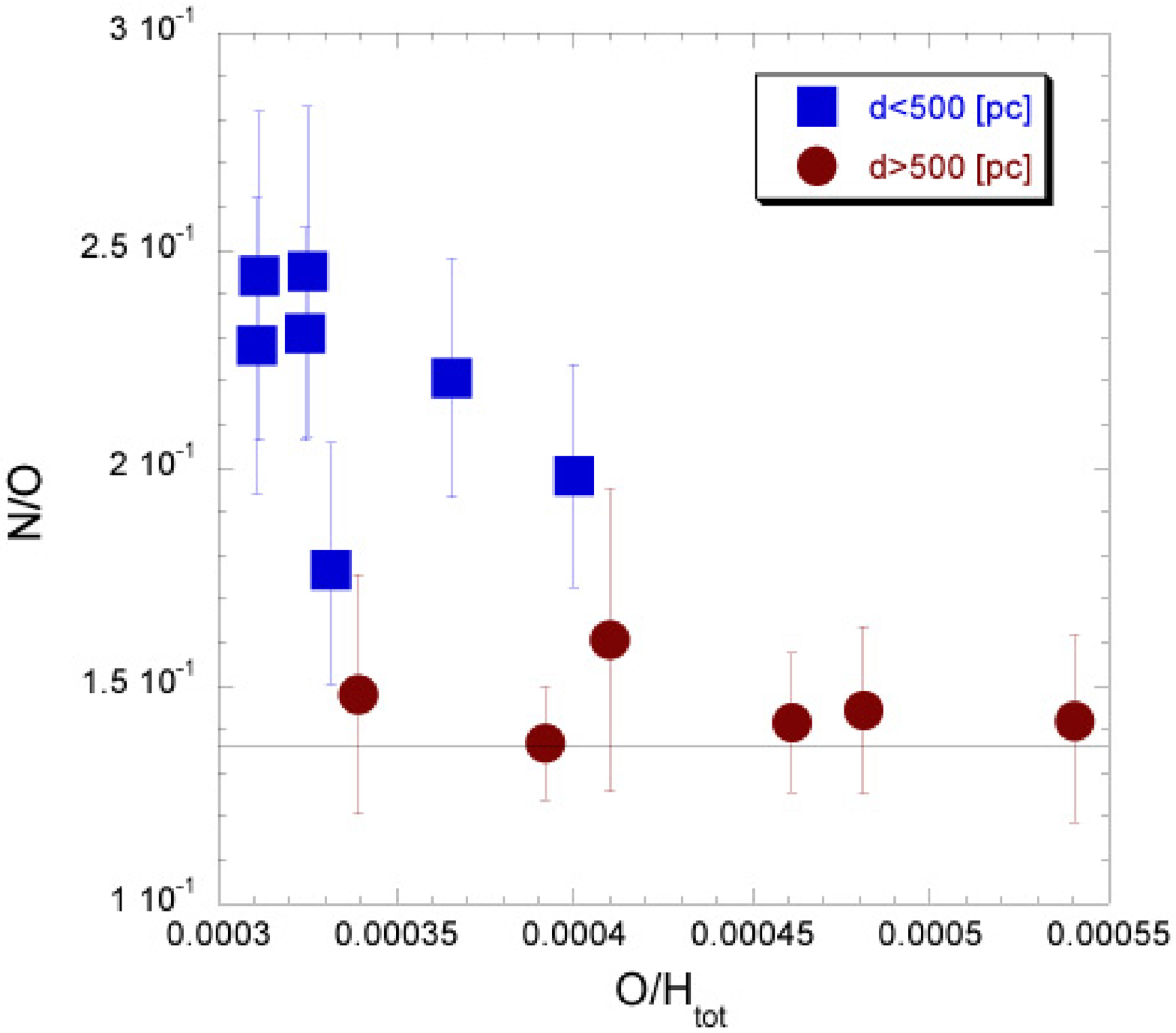}
\vspace{-0.10in}
\caption{$LEFT$ $-$ The N/O ratio versus stellar distance supports and enhances the \ion{N}{1} abundance variations reported earlier (Knauth et al. 2003).  The N/O ratio for nearby lines of sight (N/O = 0.217 $\pm$ 0.011 for $d$~$\leq$~500 pc) is $\sim$ 50\% larger than N/O for stars at further distances (N/O = 0.142 $\pm$ 0.008 for $d$~$\geq$~500 pc).  It is important to note that the lower interstellar \ion{O}{1} abundance accounts for only about a third of this difference.  This is clear evidence for enhanced nitrogen production in the Solar neighborhood or differences in interstellar mixing processes.  The thin solid line represents the Solar System ratio (N/O = 0.138$^{+0.20}_{-0.18}$; Lodders 2003).  $RIGHT$ $-$ Shows the N/O ratio compared to O/H.  The enhanced N/O at low O/H supports differences in the nucleosynthetic origin of these two species and is similar to the trend predicted by K\"{o}ppen \& Hensler (2005) if low metallicity gas has interacted with the Milky Way.}
\vspace{-0.15in}
\end{figure}
\clearpage


\begin{thebibliography}{}
\bibitem[Anders \& Grevesse(1989)]{ag89} Anders, E., \& Grevesse, N. 1989, \gca, 53, 197
\bibitem[Andr\'{e} et al.(2003)]{A03} Andr\'{e}, M., et al. 2003, \apj, 591, 1000 
\bibitem[Audouze \& Tinsley(1976)]{AT76} Audouze, J., \& Tinsley, B. M. 1976, \araa, 14, 43
\bibitem[Bohlin, Savage, \& Drake(1978)]{BSD78} Bohlin, R. C., Savage, B. D., \&
	Drake, J. F. 1978, \apj, 224, 132
\bibitem[Cartledge et al.(2001)]{C01} Cartledge, S. I. B., Meyer, D. M., \& Lauroesch, J. T. 2001, \apj, 562, 394
\bibitem[Cartledge et al.(2003)]{C03} Cartledge, S. I. B., Meyer, D. M., \& Lauroesch, J. T. 2003, \apj, 597, 408
\bibitem[Cartledge et al.(2004)]{C04} Cartledge, S. I. B., Lauroesch, J. T., Meyer, D. M., \& Sofia, U. J. 2004, \apj, 613, 1037
\bibitem[Comer\'{o}n \& Torra(1994)]{CT94} Comer\'{o}n, F., \& Torra, J. 1994, A\&A, 281, 35
\bibitem[de Avillez(2000)]{dA00} de Avillez, M. A. 2000, \mnras, 315, 479
\bibitem[Diplas \& Savage(1994)]{DS94} Diplas, A., \& Savage, B. D. 1994, 
	\apjs, 93, 211
\bibitem[Dixon, Sankrit, \& Otte(2006)]{dso06} Dixon, W. V., Sankrit, R., \& Otte, B. 2006, \apj, in press 
\bibitem[Federman, Knauth, \& Lambert(2004)]{F04} Federman, S. R., Knauth, D. C., \& Lambert, D. L. 2004, \apj, 603, L105
\bibitem[Friedman et al.(2006)]{Fried06} Friedman, S. D., H\'{e}brard, G., Tripp, T. M., Chayer, P., \& Sembach, K. R. 2006, \apj, 638, 847
\bibitem[Henry, Edmunds, \& K\"{o}ppen(2000)]{HEK00} Henry, R. B. C., Edmunds, M. G., \& K\"{o}ppen, J. 2000, \apj, 541, 660
\bibitem[Hoopes et al.(2003)]{H03} Hoopes, C. G., Sembach, K. R., H\'{e}brard, G., Moos, H. W., \& Knauth, D. C. \apj, 586, 1094
\bibitem[Jenkins et al.(2000)]{J00} Jenkins, E. B., et al. 2000, \apj, 538, L81
\bibitem[Keenan, Hibbert, \& Dufton(1985)]{KHD85} Keenan, F. P., Hibbert, A., \&
	Dufton, P. L. 1985, \aap, 147, 89
\bibitem[Knauth et al.(2003)]{K03} Knauth, D. C., Andersson, B-G, McCandliss, S. R., \& Moos, H. W. 2003, \apj, 596, L51
\bibitem[K\"{o}ppen \& Hensler(2005)]{KH05} K\"{o}ppen, J., \& Hensler, G. 2005, \aap, 434, 531
\bibitem[Lauroesch et al.(2006)]{L06} Lauroesch, J. T., Cartledge, S. I. B., Clayton, G. C., Sofia, U. J., \& Meyer, D. M. 2006, submitted
\bibitem[Lodders(2003)]{L03} Lodders, K. 2003, \apj, 591, 1220
\bibitem[Meyer et al.(1994)]{M94} Meyer, D. M., Jura, M., Hawkins, I., \& Cardelli, J. A. 1994, \apj, 437, L59
\bibitem[Meyer, Cardelli, \& Sofia(1997)]{M97} Meyer, D. M., Cardelli, J. A., \& Sofia, U. J. 1997, \apj, 490, L103
\bibitem[Meyer, Jura, \& Cardelli(1998)]{MJC98} Meyer, D. M., Jura, M., \& Cardelli, J. A. 1998, \apj, 493, 222
\bibitem[Moos et al.(2000)]{Moos00}Moos, H. W., et al. 2000, \apj, 538, L1
\bibitem[Moos et al.(2002)]{Moos02}Moos, H. W., et al. 2002, \apjs, 140, 3
\bibitem[Morton(2003)]{M03} Morton, D. C. 2003, \apjs, 149, 205
\bibitem[Morton(2004)]{M04} Morton, D. C. 2004, \apjs, 151, 403
\bibitem[Perryman et al.(1997)]{P97} Perryman, M. A. C., et al. 1997, \aap, 323, L49  
\bibitem[Rachford et al.(2002)]{R02} Rachford, B. L., et al. 2002, \apj, 577, 221
\bibitem[Sahnow et al.(2000)]{Sahnow00}Sahnow, D J., et al. 2000, \apj, 538, L7
\bibitem[Savage et al.(1977)]{S77} Savage, B. D., Drake, J. F., Budich, W., \&
	Bohlin, R. C. 1977, \apj, 216, 291
\bibitem[Savage \& Sembach(1991)]{SS91} Savage, B. D., \& Sembach, S. R. 1991, \apj, 379, 245
\bibitem[Sofia \& Jenkins(1998)]{SJ98} Sofia, U. J., \& Jenkins, E. B. 1998, \apj, 499, 951
\bibitem[Sofia, Cardelli, \& Savage(1994)]{scs94} Sofia, U. J., Cardelli, J. A., \& Savage, B. D. 1994, \apj, 430, 650
\bibitem[Sofia, Meyer, \& Cardelli(1999)]{smc99} Sofia, U. J., Meyer, D. M., \& Cardelli, J. A. 1999, \apj, 522, L137
\bibitem[Sofia(2004)]{S04a} Sofia, U. J. 2004, in ASP Conf. Ser. 309, Astrophysics of Dust, 
ed. A. N. Witt, G. C. Clayton, \& B. T. Draine (San Francisco: ASP), 393
\bibitem[Sofia et al.(2004)]{S04b} Sofia, U. J., Lauroesch, J. T., Meyer, D. M., \& Cartledge, S. I. B. 2004, \apj, 605, 272
\bibitem[Timmes, Lauroesch, \& Truran(1995)]{TLT95} Timmes, F. X., Lauroesch, J. T., \& Truran, J. W. 1995, \apj, 451, 468
\bibitem[Timmes, Woosley, \& Weaver(1995)]{TWW95} Timmes, F. X., Woosley, S. E., Weaver, T. A. 1995, \apjs, 98, 617
\bibitem[van den Hoek \& Groenewegen(1997)]{vdHG97} van den Hoek, L. B., \& Groenewegen, M. A. T. 1997, \aaps, 123, 305 
\bibitem[Welty et al.(1999)]{W99} Welty, D. E., Hobbs, L. M., Lauroesch, J. T.,
	Morton, D. C., Spitzer, L., \& York, D. G. 1999, \apjs, 124, 465
\bibitem[Wheeler, Sneden, \& Truran(1989)]{WST89} Wheeler, J. C., Sneden, C., Truran, J. W. Jr., 1989, \araa, 27, 279
\bibitem[Wood et al.(2004)]{W04} Wood, B. E., Linsky, J. L., H\'{e}brard, G., Williger, G. M., Moos, H. W., \& Blair, W. P. 2004, \apj, 609, 838


\end{thebibliography}
\end{document}